\documentclass[aps,prl,twocolumn]{revtex4-2}
\usepackage{amsmath,epsfig,mathrsfs,graphicx,amssymb,color}
\usepackage[T1]{fontenc}
\usepackage{t1enc}
\usepackage{hyperref}
\usepackage{array}
\usepackage{booktabs}

\def\be#1\ee{\begin{equation}#1\end{equation}}
\newcommand{\ba}{\begin{eqnarray} }
\newcommand{\ea}{\end{eqnarray} }

\usepackage[cp1250]{inputenc}   
\usepackage{amsmath}            
\usepackage{amsfonts}           
\usepackage{amssymb}            
\usepackage{enumerate,graphicx}
\usepackage{array}
\usepackage{booktabs}

\usepackage{physics}

\def\mb{\begin{pmatrix}}
\def\me{\end{pmatrix}}
\def\be#1\ee{\begin{equation}#1\end{equation}}

\begin{document}

\title{Quantum dimension witness with a single repeated operation}

\author{Tomasz Bia{\l}ecki$^{1}$}
\author{Tomasz Rybotycki$^{2,3,4}$}
\author{Josep Batle$^{5,6}$}
\email{jbv276@uib.es, batlequantum@gmail.com}
\author{Adam Bednorz$^{1}$}

\email{Adam.Bednorz@fuw.edu.pl}

\affiliation{$^{1}$Faculty of Physics, University of Warsaw, ul. Pasteura 5, PL02-093 Warsaw, Poland}
\affiliation{$^2$Systems Research Institute, Polish Academy of Sciences, 6 Newelska Street, PL-01-447 Warsaw, Poland}
\affiliation{$^3$Center for Theoretical Physics, Polish Academy of Sciences, Al. Lotnik{\'o}w 32/46, PL02-668 Warsaw, Poland}
\affiliation{$^4$AstroCeNT — Particle Astrophysics Science and Technology Centre — International Research Agenda,
	Nicolaus Copernicus Astronomical Center, Polish Academy of Sciences, PL-00-614 Warsaw, Poland}
\affiliation{$^5$Departament de F\'{\i}sica, Universitat de les Illes Balears,
                E-07122 Palma de Mallorca, Balearic Islands, Spain}
\affiliation{$^6$CRISP -- Centre de Recerca Independent de sa Pobla, E-07420, sa Pobla, Balearic Islands, Spain}

\begin{abstract}
We present a simple null test of a dimension of a quantum system, using a single repeated operation in the method of delays, assuming that each instance is identical and independent.
The test is well-suited to current feasible quantum technologies, with programed gates. We also analyze weaker versions of the test, assuming
unitary or almost unitary operations and derive expressions for the statistical error. The feasibility of the test is demonstrated on IBM Quantum.
The failure in one of the tested devices can indicate a lack of  identity between subsequent gates or an extra dimension in the many worlds/copies model.

\end{abstract}

\maketitle

\emph{Introduction.}
To supersede classical machines, quantum technologies must become accurate, including on-demand manipulations and error corrections.
In particular, it is important to have control over the dimension of the system, usually a qubit or its multiplicity. Otherwise,
external states lead to the deterioration of the performance, accumulated in complicated tasks.

A dimension witness is a control quantity, verifying if the system remains in the expected Hilbert space, consisting of a given (small) number 
of states. 
Its usual construction is based on the two-stage protocol, the initial preparation and final measurement \cite{gallego}, which are taken from several respective possibilities, and are independent of each other. Importantly, the preparation phase must be completed before the start of the measurement. Such early witnesses were based
on linear inequalities, tested experimentally \cite{hendr,ahr,ahr2,dim1} but they could not detect, e.g., small contributions from other states. In the latter case, it would be better to use a nonlinear witness \cite{wgp,leak}.
 A completely robust witness must be based on equality, i.e.,
a quantity, which is exactly zero up to a certain dimension(a null test).
The first such null dimension test was due to Sorkin \cite{sorkin} in the three-slit experiment \cite{tslit,btest1,btest2}  testing Born's rule \cite{born}, 
belonging to a family of precision tests of quantum mechanics, benchmarking our trust in fundamental quantum models and their actual realizations. The Sorkin
test assumes a known measurement operation whith arbitrary initial states, and therefore does not provide the information of the initial state dimension but rather
the measurement space, which was originally $3$ but can be generalized to higher values.

A more general witness test is the linear independence of the  specific outcome probability $p(y|x)$ for the preparation $x$ independent from 
the measurement $y$ by a suitable determinant \cite{dim, chen,bb22}. For a qubit (dimension 2) it requires five preparations and four measurements.
The difficulty is that it takes minimally $4\times 5\times 2=40$ logical operations (gates) and $4\times 5=20$ measurements to get a single data point. Moreover, the choice of $x$ and $y$
is a matter of luck, e.g., generated randomly \cite{opt}, because one cannot predict potential deviations.

Here, we propose a different test, using only a single operation. The initial (prepared) state can be arbitrary and the final measurement as well.
However, the operation can be repeated a given number of times, assuming that each time it remains the same. 
It can be also random (e.g. picked from a specified set), but independent from the previous choices.
The linear space spun by Toeplitz matrices
of the outcome probabilities has the rank limited by the dimension \cite{wgp,leak} (method of delays).
We construct the witness quantity as a determinant of the matrix, additionally reduced by taking preserved normalization into account. 
The test requires up to seven repetitions of the operation for a qubit, leading to $28$ logical operations and eight measurements, 
i.e. significantly fewer resources than the linear independence test.
The test can be simplified even more if one assumes that the operation is unitary or almost unitary and generalized to an arbitrary dimension.
We also find maximal deviations from zero, in higher-dimensional space. The feasibility of the test is demonstrated on IBM Quantum.
Most of tested qubits passed the test except one that showed deviations at 30 standard deviations, deserving further analysis.

\emph{Construction of witnesses.}
Our construction is based on the method of delays \cite{wgp,leak}.
The probability of the outcome determined by the measurement operator $\hat{1}\geq \hat{M}\geq 0$ after $n$ subsequent quantum operations $\mathcal E$ 
(superoperator, a linear map on an operator) 
on the initial state $\hat{P}$ (in our notation on the left, meaning the time order from the left to the right) is
\be
p_n=\mathrm{Tr}\hat{P}\mathcal E^{n}\hat{M}.\label{mrp}
\ee
The method applies to universal Markov processes (both quantum and classical), continuous and stationary in time, where $\mathcal E=e^{t\mathcal{L}^\dag}$ for the time delay $t$ and the constant Lindblad operator $\mathcal{L}$ \cite{nielsen,markov}. The operation
must preserve normalization, i.e.  $\mathcal E \hat{1}=\hat{1}$ for identity $\hat{1}$.
Suppose the linear space of possible measurements is $\leq N$, including the identity.
From the Cayley-Hamilton theorem, the characteristic polynomial $w(\mathcal E)$ is of degree $\leq N$, divisible by $\mathcal{E}-1$ since one of the eigenvalues is $1$.
We construct a $N\times N$-dimensional Toeplitz square matrix $W_N$
with entries
\be
W_{N,jk}=p_{j+k}-p_{j+k+1}\label{wit1}
\ee
for $j,k=0\dots N-1$. Then in our case $\det W_N=0$. Alternatively $\det W_N=\det \tilde{W}_N$ with the $N+1\times N+1$ matrix
$\tilde{W}_{N,j,k}=p_{j+k}$ for $j\leq N$ and $\tilde{W}_{N,N+1\,k}=1$.

A quantum operation  (completely positive map) $\mathcal E$ has a Kraus decomposition
 $\mathcal E\hat{M}=\sum_j\hat{K}_j\hat{M}\hat{K}_j^\dag$, $\sum_j \hat{K}_j\hat{K}_j^\dag=\hat{1}$.
For unitary operations, the sum consists of a single, unitary $\hat{K}$. A general $d$-dimensional state can be written in terms of $d^2-1$ traceless
Gell-Mann matrices (Pauli matrices $\sigma_x$ and $\sigma_y$ for each pair of indices and $\sigma_z$ for pairs $1,n$, with $n=2\dots d$), 
and the trace component $\hat{1}/d$. Restricting to real space, we get only half of the off-diagonal Gell-Mann matrices, giving the dimension
$d(d+1)/2$ (including trace). The operation $\mathcal{E}$ has then maximally $d^2$ eigenvalues, with one of them equal to $1$ for trace preservation.
All non-real eigenvalues must appear in conjugate pairs as $\mathcal E$ also preserves Hermiticity.
Then $\det W_N=0$ for:  $N\geq d$ classically,  $N\geq d(d+1)/2$ in real quantum space, and $N\geq d^2$ in complex quantum space,
because the number of linearly independent columns is limited by the dimension of the space of accessible operations.
A list of the classical maxima of Eq. (\ref{wit1}) and corresponding probabilities is given in Table \ref{tmax}.
 In the unitary case, the eigenvalues of $\hat{K}$ have $|\lambda_j|=1$ and the eigenvalues of the adjoint operator read
$\lambda_j\lambda^\ast_k$ for $j,k=1\dots d$.
In the simplest quantum (either real or complex) case, $d=2$ we have eigenvalues, $1$ (twice), and $e^{\pm i\phi}$.
Then, additionally for all pairs $\alpha,\beta$ in this set, we have
\be
(\alpha-\beta)^2(\alpha\beta-1)(\alpha-1)(\beta-1)=0.\label{abb}
\ee
Therefore, expanding $p_n=\sum_jA_j\lambda_j^n$,
\begin{equation}
F_1=p_1^2 + p_0 (p_3-p_2) - p_2 (p_1-p_3) - p_3^2 + (p_2-p_1) p_4 \label{wit2}
\end{equation}
must vanish, as it contains Eq. (\ref{abb}) for $\alpha,\beta=\lambda_j$. Witnesses for $d>2$ are analogous but lengthy due to various combinations of eigenvalues.
Another interesting case is when $\mathcal E$ is almost unitary, meaning that $|\lambda_j|$ are close to 1. To discriminate a higher dimension from
minimal non-unitary effects (e.g., decoherence, relaxation) for $d=2$, we use a modification of Eq. (\ref{abb})
\be
[(\alpha-\beta)(\alpha\beta-1)(\alpha-1)(\beta-1)]^2=0,\label{aab}
\ee
giving analogously the witness
\begin{align}
&F_2=\nonumber\\
&(p_2-2p_3+p_4)(p_6-2p_3+p_0) - (p_2-p_1+p_4-p_5)^2.\label{wit3}
\end{align}
If $1-|\lambda_j|\sim \epsilon$, then Eq. (\ref{aab}) is $\sim\epsilon^2$, in contrast to Eq. (\ref{abb}) $\sim\epsilon$.
In particular, taking $\hat{P}=\hat{M}=|0\rangle\langle 0|$ and $\lambda=1,\pm i(1-\epsilon)$, we get  Eq. (\ref{wit2}) equal to $8\epsilon$ while
Eq. (\ref{wit3}) is equal to $32\epsilon^2$. Practical relaxation, depolarization, and phase damping lead to non-unitarities of the order $10^{-4}$
in the existing implementations, so this kind of test, although less general than Eq. (\ref{wit1}), can verify the dimension with fewer resources.

A common problem of quantum manipulations is leakage to external states, which no longer participate in the dynamics but still affect the measurement. This
case can be resolved by adding an extra classical (sink) state so that the dimension is shifted by $1$, i.e. 
$\det W_N=0$ for $N>d^2$ in complex and $N>d(d+1)/2$ in real space. 

\begin{table}
\begin{tabular}{*{3}{c}}
\toprule
$N$&$W_N$&$p_{0,\dots, 2N-1}$\\
\midrule
$1$&$1$&$0,1$\\
$2$&$-1$&$1,1,0,0$\\
$3$&$1.25$&$1,0,0,1,0.5,1$\\
$4$&$2.088$&$0,1,0,0,1,\sqrt{2/3},1,0$\\
$5$&$4$&$1,0,1,1,1,0,0,1,0,1$\\
$6$&$-8$&$0,0,0,1,1,0, 1, 0, 0, 0, 1,1$\\
$7$&$16$&$1,1,0,1,0,0, 0, 1, 1,0, 1, 0, 1,1$\\
$8$&$18$&$0,1,0,0,1,1,1,0,1,0,1,0,0,1,1,0$\\
$9$&$64$&$0,1,0,0,1,0,1,1,1,0,0,0,0,1,0,0,1,0$\\
\bottomrule
\end{tabular}
\caption{Classical maxima of $W_N$ given by Eq. (\ref{wit1}) for initial values of $N$ with corresponding lists of all probabilities that it depends on.
$W_4=1+4\sqrt{6}/9$.}
\label{tmax}
\end{table}

\emph{Error analysis}.
The dominant source of errors is finite statistics. For any witness, which is a function of binary probabilities
in independent experiments, $F(\{p_j\})$ (here $F=F_1,F_2,W_N$), we can distinguish $\tilde{p}_j=n_j/N_j$, the actual frequency of $n_j$ successes (e.g. result $1$)
for $N_j$ repetitions. In contrast, $p_j$ is the limit at $N_j\to \infty$. In the case of randomly chosen $j$, we can assume independence between subsequent experiments, so
\begin{equation}
\langle \delta p_j\rangle=0,\;\langle\delta p_j\delta p_k\rangle=\frac{p_j(1-p_j)}{N_j}\delta_{jk}
\end{equation}
for $\delta p_j=\tilde{p}_j-p_j$.
At large $N_j$ we can expand
\begin{equation}
\delta F\simeq\sum_j\frac{\partial F}{\partial p_j}\delta p_j+\sum_{jk}
\frac{\partial^2 F}{2\partial p_j\partial p_k}\delta p_j\delta p_k
\end{equation}
for $\delta F=F(\{\tilde{p}_j\})-F(\{p_j\})$, which gives the dominant shift of the average and variance
\begin{align}
&\langle \delta F\rangle\simeq\sum_j\frac{\partial^2 F}{2\partial p_j^2}b_j/N_j,\nonumber\\
&\langle (\delta F)^2\rangle\simeq\sum_j\left(\frac{\partial F}{\partial p_j}\right)^2 b_j/N_j,\label{shift}
\end{align}
denoting $b_j=p_j(1-p_j)$.
In the limit $N_j\to \infty$ the variance dominates. Setting equal $N_j=N$ we can find the error for our witnesses,
using the derivative of actually measured $F(\{\tilde{p}_j\})$ in place of $F(\{p_j\})$, assuming they are close to each other.
In particular for Eq. (\ref{wit1})
the variance reads
\begin{equation}
\langle(\delta F)^2\rangle\simeq \sum_{i}b_i\left(\sum_j(\mathrm{Adj}\: W_{j,i-j}-\mathrm{Adj}\: W_{j-1,i-j})\right)^2
\end{equation}
where Adj is the adjoint matrix (matrix of minors of $W$, with a given row and column croddes out, and then transposed). Here the entries are $0$
outside of the size. Note that the identity 
$W^{-1}\det W=\mathrm{Adj}W$ makes no sense here as $\det W=0$ in the limit $N\to\infty$.
For
Eq. (\ref{wit2}), we have
\begin{align}
&\langle(\delta F)^2\rangle\simeq b_0(p_3-p_2)^2+b_1(2p_1-p_2-p_4)^2\nonumber\\
&+b_2(p_0+p_1-p_4)^2+b_3(2p_3-p_2-p_0)^2+b_4(p_1-p_2)^2,
\end{align}
and for Eq. (\ref{wit3}),
\begin{align}
&\langle(\delta F)^2\rangle\simeq (b_0+b_6)(p_2-2p_3+p_4)^2\nonumber\\
&+(b_1+b_5)(p_2-p_1+p_4-p_5)^2\nonumber\\
&+(b_2+b_4)(2p_2-2p_1+2p_4-2p_5-p_6+2p_3-p_0)^2\nonumber\\
&+4b_3(p_2+p_6-4p_3+p_0+p_4)^2.
\end{align}

In the case of small $\partial F/\partial p_j$, which happens in the degenerate case, when e.g. the operations are restricted near the real subspace, there is a competitive contribution to $\langle(\delta F)^2\rangle$,
\begin{equation}
\sum_{ij}\left(\frac{\partial^2 F}{\partial p_i\partial p_j}\right)^2\frac{b_ib_j}{2N_iN_j},\label{dext}
\end{equation}
which reads for (\ref{wit3}),
\begin{align}
&[(b_0+2b_3+b_6)(b_2+2b_3+b_4)/2+2b_3^2+\nonumber\\
&(b_1+b_2+b_4+b_5)^2+b^2_1+b^2_2+b^2_4+b^2_5]/N^2.
\end{align}

In principle, one should also consider the cross term
\begin{equation}
\sum_i \frac{\partial^2 F}{\partial p_i^2}\frac{\partial F}{\partial p_i}t_i/N_i^2,
\end{equation}
where $t_i=p_i(1-p_i)(1-2p_i)$ is the third cumulant. By $2ab\leq a^2+b^2$ inequality applied to
\begin{align}
&a=\frac{\partial F}{\partial p_i}\sqrt{p_i(1-p_i)}/N_i^{1/2},\nonumber\\
&b=\frac{\partial^2 F}{\partial p^2_i}\sqrt{p_i(1-p_i)}(1-2p_i)/N_i,
\end{align}
we see that it is by $\sim N_i^{-1/2}$ smaller than the previous terms in the limit of large $N_j$.

\begin{figure}
\includegraphics[scale=.7] {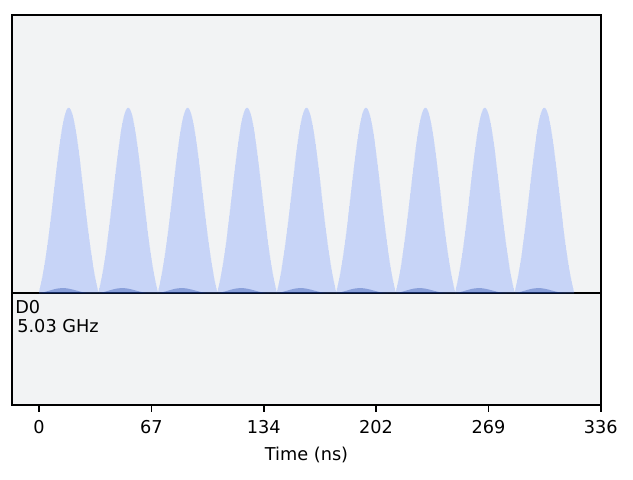}
\caption{The actual waveform for the gate pulse (here lima qubit $0$), here with $k=9$ gates, $\pi/2$ rotations about the $x$ axis, Eq. (\ref{smat}).
The two initial gates are for preparation and so $k$ is from $2$ to $9$ to test all the witnesses for $d=2$.}
\label{wav}
\end{figure}

\emph{Demonstration on IBM Quantum}.
We have tested the witness (\ref{wit1}) for $N=3,4$ and (\ref{wit3}) for qubits on IBM Quantum.
The initial state $|1\rangle$ undergoes $k$ rotations $\pi/2$ about the $X$ direction, i.e.
\begin{equation}
\hat{S}=\sqrt{X}=\frac{1}{\sqrt{2}}\begin{pmatrix}
1&-i\\
-i&1\end{pmatrix}\label{smat},
\end{equation}
in the basis $|0\rangle,|1\rangle$, and finally measures the state $|1\rangle$. In the ideal case, it is a perfect unitary operation with eigenvalues $\pm i$ and $1$.
Then $W_3=W_4=F_1=F_2=0$. However, due to decoherence and decay, this is inexact, leading to the tolerance at the level $\sim 10^{-6}$ for $F_2$
assuming a decay rate $\sim 10^{-3}$. In principle, the decay should not affect $W_3$ as the quantum operation is restricted to the Bloch
great circle (with the interior) and the operation (gate) should have a calibrated phase. Nevertheless, this assumption may be still not perfect
and so only $W_4$ is the full test to the qubit space.

\begin{figure}
\includegraphics[scale=.7]{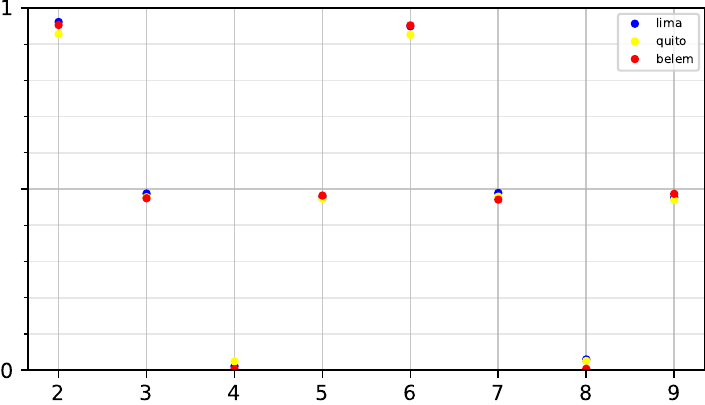}
\caption{Averaged probabilities of the measurement  of the states $|1\rangle$ for $k=2\dots 9$ gates in the case of our demonstrations.
The data show relatively faithful oscillations, with small variations to be checked by the witnesses.}
\label{prob}
\end{figure}

\begin{table*}
\begin{tabular}{*{6}{c}}
\toprule
device&T1 (microsecond)&T2 (microsecond)& frequency (GHz)& anharmonicity (GHz)&gate error\\
\midrule
lima&117&169&5.03&-0.34&$5\cdot 10^{-4}$\\
quito&16&38&5.30&-0.33&$1.8\cdot 10^{-3}$\\
belem&118&122&5.09&-0.34&$1.8\cdot 10^{-4}$\\
\bottomrule
\end{tabular}
\caption{The characteristics of the qubits used in the demonstration, times T1 (relaxation) and T2 (coherence),
frequency between the 0 and 1 level, anharmonicity (frequency between 1 and 2 levels above 0-1 transition), and error of the gate $S=\sqrt{X}$ used in the test. The duration of the single gate pulse is always 35ns.}
\label{tech}
\end{table*}
\begin{table*}
\begin{tabular}{*{6}{c}}
\toprule
device&witness&$F$&$F-\Delta F$&$\sigma_F$&$\sigma'_F$\\
\midrule
&$F_2$&$-1.054\cdot 10^{-4}$& $-1.049\cdot 10^{-4}$&$7.329\cdot 10^{-6}$&$7.325\cdot 10^{-6}$\\
lima&$W_3$&$1.018 \cdot 10^{-4}$&$1.023 \cdot 10^{-4}$&$7.469146 \cdot 10^{-5}$&$7.469140 \cdot 10^{-5}$\\
&$W_4$&$2.65\cdot 10^{-5}$&$2.54 \cdot 10^{-5}$&$7.6\cdot 10^{-7}$&  $7.4\cdot 10^{-7}$\\
\midrule
&$F_2$&$-1.38\cdot 10^{-5}$&$-1.30 \cdot 10^{-5}$&$5.094\cdot 10^{-6}$&$5.086\cdot 10^{-6}$\\
quito&$W_3$&$1.895 \cdot 10^{-4}$&$1.899\cdot 10^{-4}$&$8.63552 \cdot 10^{-5}$&$8.63551\cdot 10^{-5}$\\
&$W_4$&$1.2\cdot 10^{-6}$&$1.5\cdot 10^{-7}$&$4.0\cdot 10^{-7}$&$3.5\cdot 10^{-7}$\\
\midrule
&$F_2$&$6.08\cdot 10^{-5}$&$6.12\cdot 10^{-5}$&$6.935\cdot 10^{-6}$&$6.931\cdot 10^{-6}$\\
belem&$W_3$&$-3.837\cdot 10^{-4}$&$-3.831\cdot 10^{-4}$&$8.20748\cdot 10^{-5}$&$8.20747\cdot 10^{-5}$\\
&$W_4$&$1.6\cdot 10^{-6}$&$5.04 \cdot 10^{-7}$&$3.7\cdot 10^{-7}$&$3.2\cdot 10^{-7}$\\
\bottomrule
\end{tabular}
\caption{The values to the witnesses inferred from the demonstration on IBM Quantum. Here, $\Delta F$ is the shift according to Eq. (\ref{shift}),
$\sigma_F$ is the error with both Eqs. (\ref{shift}) and (\ref{dext}) taken into account, and $\sigma'_F$ is the error taking only
(\ref{shift}) for comparison.}
\label{tab2}
\end{table*}

We have probed our witnesses on three devices,
lima, quito, and belem, qubit $0$, applying $k$  gates $\hat{S}$ to the ground state $|0\rangle$. Two gates prepare the initial state $|1\rangle$ so the test requires $k=2\dots 9$. Their technical characterization is listed in Table \ref{tech}.
The test consisted of circuits with randomly shuffled $k$, with $239$, $160$, and $200$ jobs, respectively (see Fig. \ref{wav} for the pulse realization). Each job contains $12$
repetitions of the circuit due to the limit of $100$ circuits and $20000$ shots (repetitions of each circuit). 
The probability roughly repeats $1$, $0.5$, $0$, $0.5$, as the gates rotate between the basis states with the middle superposition (see Fig. \ref{prob}).
Due to calibration drifts
and nonlinearity of our witnesses we decided to calculate the witness for each job and then average them, averaging the variances and shifts, too.
Since the first order errors of $W_4$ and $F_2$ can be very small, we have checked also the 
higher order contribution, Eq. (\ref{dext}), and the shift of the average, Eq. (\ref{shift}). It turned out that the shift is about $1\%$ of the observed value
while the dominant error comes from Eq. (\ref{shift}). We have collected these results with errors in Table \ref{tab2}.

Only in the case of lima, we see $W_4$ more than $30$ standard deviations from $0$ indicating failure of the assumed dimension or identity between subsequent gates. For other devices, the values are apparently moderate but after shift correction they are within the error. The standard error from Eq. (\ref{shift}) is dominating the next correction, Eq. (\ref{dext}). Other witnesses also indicate
a nonzero value, but these can be in principle due to some technical imperfections.

\begin{figure}
\includegraphics[scale=.7]{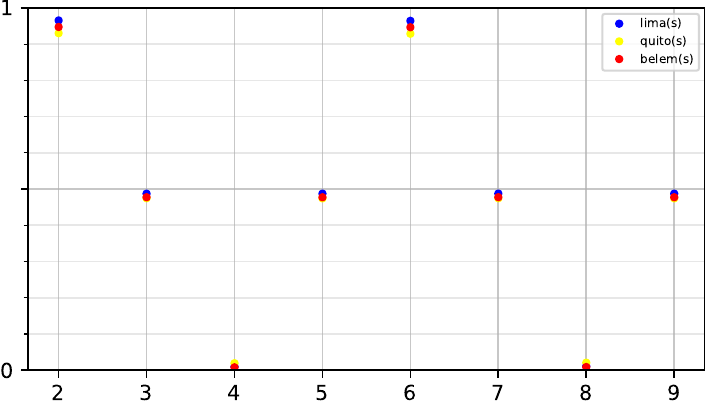}
\caption{Averaged probabilities of the measurement  of the states $|1\rangle$ for $k=2\dots 9$ gates in the case of simulations, compared to Fig. \ref{prob}}
\label{probap}
\end{figure}

We have also checked the simulated experiments using the noise models from lima, quito, and belem, for the same number of jobs as in real experiments 
(see Fig. \ref{probap}).
All the witnesses remain zero up to the experimental error, as shown in Table \ref{tabap}. A moderate exception is $F_2$ for belem, about 3 standard deviations.
We believe it can be attributed to second-order deviations from the simulated decay rate.
 The data and scripts are available in the open repository \cite{zen}.

\begin{table*}
\begin{tabular}{*{6}{c}}
\toprule
device&witness&$F$&$F-\Delta F$&$\sigma_F$&$\sigma'_F$\\
\midrule
&$F_2$&$1.3\cdot 10^{-7}$& $4.9\cdot 10^{-7}$&$6.0\cdot 10^{-7}$&$5.6\cdot 10^{-7}$\\
lima(s)&$W_3$&$4.86 \cdot 10^{-5}$&$4.91\cdot 10^{-5}$&$7.626514 \cdot 10^{-5}$&$7.626508 \cdot 10^{-5}$\\
&$W_4$&$1.1\cdot 10^{-6}$&$3.7 \cdot 10^{-8}$&$2.9\cdot 10^{-7}$&  $2.4\cdot 10^{-7}$\\
\midrule
&$F_2$&$-1.34\cdot 10^{-6}$&$-6.3 \cdot 10^{-7}$&$8.9\cdot 10^{-7}$&$8.4\cdot 10^{-7}$\\
quito(s)&$W_3$&$-2.00 \cdot 10^{-5}$&$-1.95\cdot 10^{-5}$&$8.71329 \cdot 10^{-5}$&$8.71328\cdot 10^{-5}$\\
&$W_4$&$9.5\cdot 10^{-7}$&$-1.4\cdot 10^{-7}$&$3.4\cdot 10^{-7}$&$2.8\cdot 10^{-7}$\\
\midrule
&$F_2$&$1.8\cdot 10^{-6}$&$2.3\cdot 10^{-6}$&$7.6\cdot 10^{-7}$&$7.2\cdot 10^{-7}$\\
belem(s)&$W_3$&$-2.36\cdot 10^{-5}$&$-2.31\cdot 10^{-5}$&$8.117147\cdot 10^{-5}$&$8.117140\cdot 10^{-5}$\\
&$W_4$&$1.3\cdot 10^{-6}$&$1.9 \cdot 10^{-7}$&$3.3\cdot 10^{-7}$&$2.8\cdot 10^{-7}$\\
\bottomrule
\end{tabular}
\caption{The values to the witnesses from simulations, with the same number of jobs and notation as in Table \ref{tab2}.}
\label{tabap}
\end{table*}

\begin{figure}
\includegraphics[scale=.7]{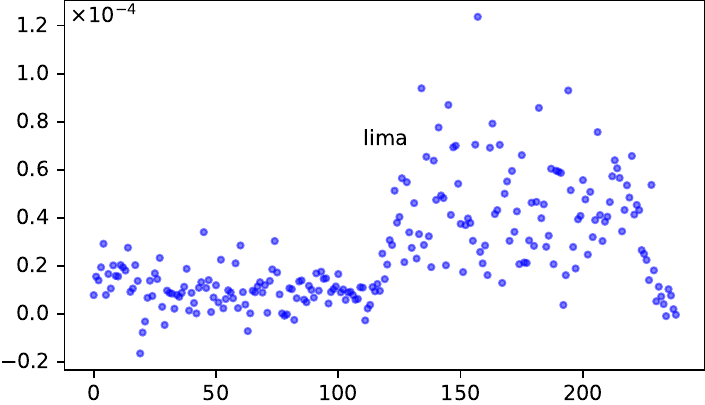}
\includegraphics[scale=.7]{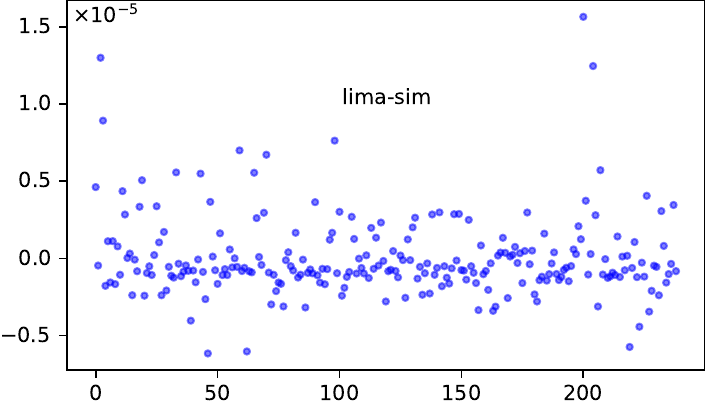}
\caption{The witness $W_4-\Delta W_4$ for individual jobs (indexed) for lima in the real experiment (top) and simulations (bottom). Note the drift in the real experiment.}
\label{limaj}
\end{figure}

To investigate a large deviation of $W_4$ for lima, we have checked the witnesses for the individual jobs. See Fig. \ref{limaj}. One can notice that the deviation
changes with the job index, indicating some calibration drift. No such effect is observed in simulations.

\emph{Discussion}.
The presented test of dimension using a single repeated operation turns out to be a reliable diagnostic tool to check the dimension of a qubit
and other finite systems. The advantage of the test is its simplicity, not demanding the specific form of the quantum operation, just its dimension, requiring minimal resources in terms of the number of operations and parameters, fewer than in Refs. \cite{dim,opt}. Although we test the zero of the special determinant as
 in Refs. \cite{dim, chen,bb22}, we rely on different assumptions making  the test complementary to the previous proposals. The observed significant deviations need further investigation of their cause. It may be just technical due to unspecified transitions to other states due to anharmonicity (Table \ref{tech}), a transition to higher excited state is unlikely) or
fundamental due to states beyond simple models predicting extra dimensions, as many worlds/copies \cite{plaga,abadp}. The tests can be also further developed in various directions, higher dimensions, 
entangled states, or combined operations. We believe that the test can be conducted also on other implementations of qubits such as photons and ion traps.

\emph{Acknowledgments}. The results have been created using IBM Quantum. The views expressed are those of the authors and do not reflect the official policy or position of the IBM Quantum team. We thank Jakub Tworzyd{\l}o for advice, technical support, and discussions, and Witold Bednorz for consultations on error analysis.
TR acknowledges the financial support by TEAM-NET project co-financed by EU within the Smart Growth Operational Programme 
(Contract No.  POIR.04.04.00-00-17C1/18-00).

\end{document}